\documentclass[%
 aip,
 amsmath,amssymb,
 reprint,%
]{revtex4-1}

\usepackage{graphicx}
\usepackage{dcolumn}
\usepackage{bm}

\usepackage{algpseudocode}
\usepackage{float}
\usepackage[lined,ruled]{algorithm2e}

\usepackage[utf8]{inputenc}
\usepackage[T1]{fontenc}
\usepackage{mathptmx}
\usepackage{etoolbox}
\usepackage{xcolor}

\makeatletter
\def\@email#1#2{%
 \endgroup
 \patchcmd{\titleblock@produce}
  {\frontmatter@RRAPformat}
  {\frontmatter@RRAPformat{\produce@RRAP{*#1\href{mailto:#2}{#2}}}\frontmatter@RRAPformat}
  {}{}
}%
\makeatother
\begin{document}

\preprint{AIP/123-QED}

\title[Influence Maximization in Hypergraphs]{Influence Maximization based on Threshold Models in Hypergraphs}
\author{Renquan Zhang}
 \affiliation{ 
School of Mathematical Sciences, Dalian University of Technology
}%
\author{Xilong Qu}%
 \affiliation{ 
School of Mathematical Sciences, Dalian University of Technology
}%

\author{Qiang Zhang}
\affiliation{%
School of Computer Science and Technology, Dalian University of Technology
}%

\author{Xirong Xu}
\affiliation{%
School of Computer Science and Technology, Dalian University of Technology
}%

\author{Sen Pei*}
 \email{sp3449@cumc.columbia.edu.}
\affiliation{%
Department of Environmental Health Sciences, Mailman School of Public Health, Columbia University
}%

\date{\today}

\begin{abstract}
Influence Maximization problem has received significant attention in recent years due to its application in various domains such as product recommendation, public opinion dissemination, and disease propagation. This paper proposes a theoretical analysis framework for collective influence in hypergraphs, focusing on identifying a set of seeds that maximize influence in threshold models. Firstly, we extend the Message Passing method from pairwise networks to hypergraphs to accurately describe the activation process in threshold models. Then we introduce the concept of hypergraph collective influence (HCI) to measure the influence of nodes. Subsequently, We design an algorithm, HCI-TM, to select the Influence Maximization Set, taking into account both node and hyperedge activation. Numerical simulations demonstrate that HCI-TM outperforms several competing algorithms in synthetic and real-world hypergraphs. Furthermore, we find that HCI can be used as a tool to predict the occurrence of cascading phenomena. Notably, we find that HCI-TM algorithm works better for larger average hyperdegrees in Erdős-Rényi (ER) hypergraphs and smaller power-law exponents in scale-free (SF) hypergraphs. 
\end{abstract}

\maketitle

\begin{quotation}
The innovation of this study lies in several key aspects. Firstly, we extend the Message Passing equation to threshold models in hypergraphs, providing a more comprehensive and accurate understanding of the activation process. Secondly, we introduce the concept of Hypergraph Collective Influence (HCI), which quantifies the collective influence of nodes in hypergraphs. HCI offers a novel approach to measuring the influence potential of individual nodes in hypergraphs. Thirdly, we propose the HCI-TM algorithm, designed to select the Influence Maximization Set in hypergraphs based on threshold models. Notably, the HCI-TM algorithm considers both individual node influence and the collective impact of hyperedges, resulting in more effective Influence Maximization strategies in hypergraphs. Through extensive simulations on synthetic and real-world hypergraphs, we demonstrate that the HCI-TM algorithm outperforms other classical algorithms. Additionally, HCI proves useful in predicting cascading phenomena. Furthermore, our simulations reveal interesting observations, indicating that the HCI-TM algorithm performs better with larger average hyperdegrees in ER hypergraphs and smaller power-law exponents in SF hypergraphs. These findings provide valuable insights into the influence of hypergraph characteristics on HCI-TM algorithm performance.
\end{quotation}

\section{Introduction:}

The Influence Maximization problem in complex networks, which aims to select a fixed number of seed nodes to maximize the scope of spreading, is widely recognized as an NP-hard problem \cite{Granovetter1978Threshold,2003Maximizing}. The critical role played by a small subset of nodes in the spreading process has been widely acknowledged in various applications\cite{1999Accelerating,2010Influencing}. For instance, a limited number of individuals possess significant influence and can shape public opinion, thereby facilitating the transmission of ideas. Similarly, identifying and immunizing super spreaders in disease transmission networks is crucial for effectively containing outbreaks. Due to its broad applications in disease control \cite{2010Scalable,ZHANG2023133677}, marketing strategy\cite{2019Community}, and other fields\cite{2015Online}, this problem has emerged as a prominent topic in recent academic research.

In pairwise networks, the topic of Influence Maximation has been extensively researched\cite{2020Influencer,2014Searching}, and a large number of heuristic algorithms, such as Betweenness\cite{Linton1978Centrality}, High Degree (HD)\cite{2000Error}, PageRank\cite{1998The}, K-core\cite{1983Network}, and CI\cite{Morone2015Influence,2016Efficient,2018Dynamic}, have been presented. However, with the continuous development of complex system modeling \cite{2006Diversity}, researchers are increasingly realizing that many real-world interactions involve multiple individuals. Pairwise networks are difficult to distinguish and describe high-order interactions.  For example, a research group can consist of multiple members, a WeChat group can have multiple participants, and a tweet can receive multiple comments. Hypergraphs provide a modeling framework that can distinguish and represent such high-order interactions\cite{2020Networks,2005Beyond}. In recent years, with the growing development of hypergraph theory and its wide-ranging applications, there has been a greater interest in Influence Maximization problem in hypergraphs. Xie et al.\cite{2023HDA} proposed an effective adaptive heuristic algorithm to find the optimal set of influence seeds in hypergraphs. Zhu et al.\cite{2021Social} investigated how to maximize social influence in social hypergraphs. Austin R. et al.\cite{2019Three} extended the feature vector center from graphs to hypergraphs. These\cite{2023HDA,2021Social,2019Three,10031164,2021Social,unknown} are important progress on Influence Maximation problem in hypergraphs. However, it remains a challenging problem, as many existing heuristic algorithms lack sufficient mathematical  analysis and overlook the dependence of nodes' neighbors. It is widely recognized that the Message Passing equation\cite{PhysRevE.82.016101} has proven effective in reducing loops in pairwise networks by removing nodes, enhancing node independence, and providing a more precise definition of the information transmission process. In this study, we aim to extend the existing theory to hypergraphs in order to investigate the Influence Maximization problem. Our main contribution lies in the proposal of a novel Message Passing equation tailored specifically for threshold models in hypergraphs, thereby providing a more effective framework for analyzing and optimizing influence propagation in hypergraphs.

Linear Threshold Models (LTM) can describe a large variety of real-world phenomena, such as disease spread\cite{Cheng2020}, public opinion diffusion\cite{2013Influence}, information dissemination\cite{2014Message}, and behavior adoption\cite{0The,Granovetter1978Threshold,2011Creating}. However, there haven't been many studies\cite{2022threshold} on threshold models in hypergraphs up to now. Therefore, we aim to investigate the optimal influence of nodes based on the threshold models in hypergraphs. The main contributions of this paper are as follows:

\begin{itemize}
\item In this paper, we extend the Message Passing Method from pairwise networks to hypergraphs in order to reduce the dependence of nodes' neighbors. Then we analyze the self-satisfying equation to obtain the Hypergraph Collective Influence (HCI), which serves as a metric for quantifying the collective influence of nodes in hypergraphs.

\item To select the optimal influence node in hypergraphs, we propose the HCI-TM algorithm. Compared to other methods, our algorithm not only maximizes node activation scale but also hyperedge activation scale.

\item Through numerical simulations, we demonstrate that the HCI-TM algorithm outperforms other classical algorithms in both synthetic hypergraphs and real-world hypergraphs. Additionally, we find that HCI can be used as a basis for assessing the occurrence of cascade phenomena.

\item Our numerical simulations further reveal that  larger average hyperdegrees in ER hypergraphs and smaller power-law exponents in SF hypergraphs could enhance the performance of HCI-TM algorithm.

\end{itemize}

The article structure is as follows: Section 1 provides the background, research motivation, and innovative aspects of our study. In Section 2, we introduce the threshold models in hypergraphs. Section 3 focuses on extending the Message Passing Method from pairwise networks to hypergraphs. We propose the concept of HCI and design the HCI-TM algorithm to select the Influence Maximization Set based on threshold models in hypergraphs. Section 4 presents the validation of the HCI-TM algorithm through numerical simulations conducted on both synthetic and real-world hypergraphs. The results demonstrate the superiority and robustness of HCI-TM in diverse hypergraph structures. And we also find that HCI can be used as a tool to predict the occurrence of cascading phenomen. Finally, in Section 5, we summarize our findings and provide insightful suggestions for future research directions.

\section{Threshold Models in Hypergraphs:}

\begin{figure}
	\centering
		\includegraphics[scale=.5]{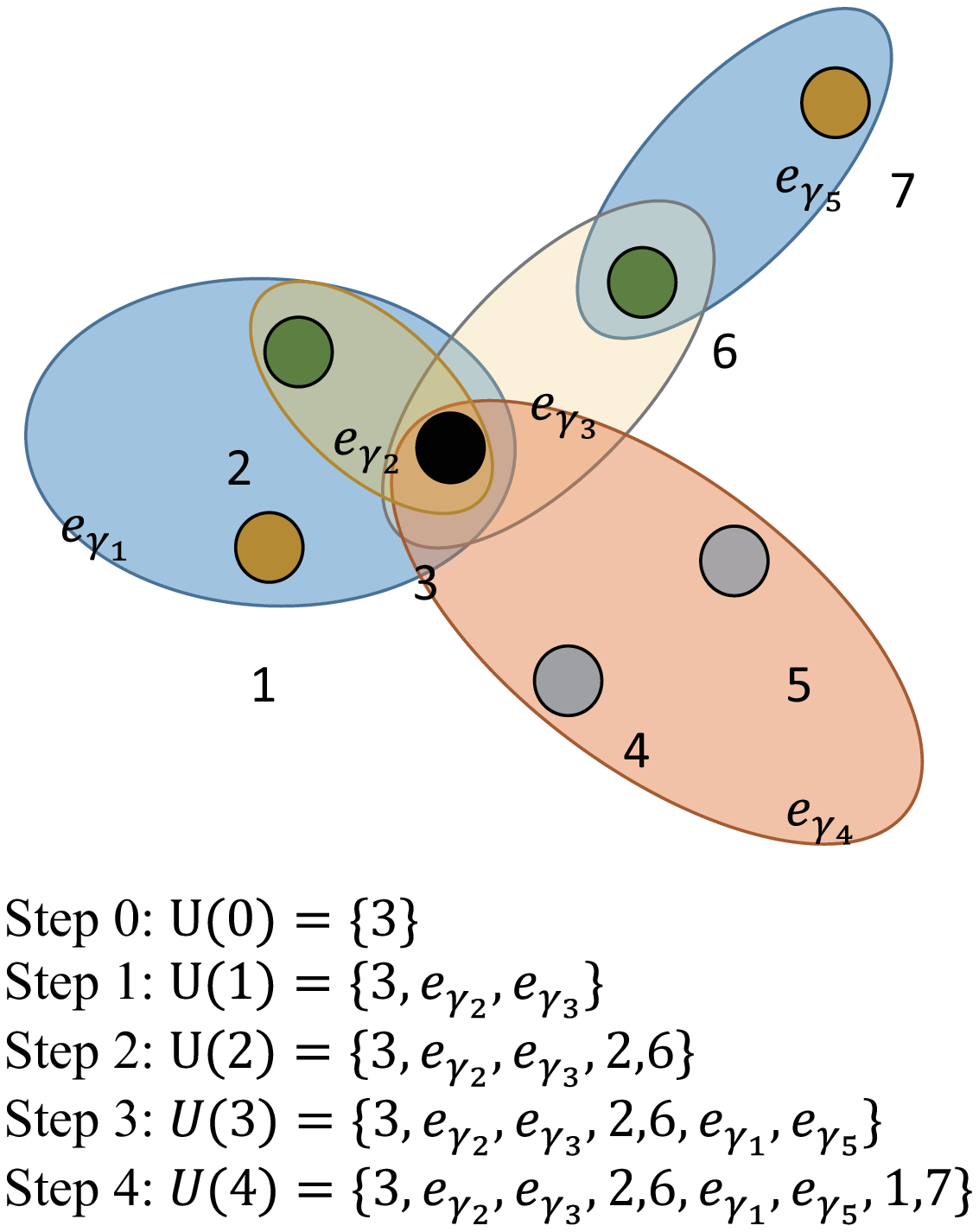}
	\caption{The diagram of the hypergraph spread process depicts nodes as circles and hyperedges as ovals. Each hyperedge comprises multiple nodes, and each node is associated with multiple hyperedges. The colors assigned to the hyperedges and nodes represent the order of activation.}
	\label{FIG:1}
\end{figure}

A hypergraph is a generalization of a graph where a hyperedge can connect any number of nodes, unlike regular graphs which only allow connections between two nodes. Hypergraph $\mathbb{H}=(V,E)$ is a subset system of finite sets, where $V$ is the set of nodes in the hypergraph, and $E$ is the set of hyperedges in the hypergraph. In a hypergraph with $N$ nodes and $M$ hyperedges, the topology structure can be represented by the correlation matrix $\{H_{ie_\gamma}\}_{N\times M}$. Here, $H_{ie_\gamma}$ is equal to 1 if node $i$ is associated with hyperedge $e_{\gamma}$, and 0 otherwise. We use vector $ \textbf{n}=(n_1,n_2,...,n_N)^{T}$ to indicate whether node $i$ is a seed node. If node $i$ is a seed node, the $ith$ component $n_i=1$, otherwise, $n_i=0$. So the fraction of seed nodes $q = \sum\limits_{i = 1}^N {\frac{{{n_i}}}{N}}$.
During the activating process, the states of each node and hyperedge can be either active or inactive, depending on the rule of threshold models in hypergraphs. Assuming that hyperedge $e_\gamma$ consists of $N_\gamma$ nodes and $m_\gamma$ 
 nodes of them are in the state of active. The hyperedge $e_\gamma$ becomes activated if the fraction of activated nodes $\frac{m_\gamma}{N_\gamma}$ reached or exceeded the threshold value $t_\gamma \in (0,1)$. Once a hyperedge becomes activated, all the nodes within that hyperedge will also become active in the next step of the propagation process. This process continues until there are no more newly activated nodes or hyperedges. To illustrate this concept, we can refer to Fig.(\ref{FIG:1}), which shows an example of the activation process in a hypergraph. In this example, all the threshold values of the hyperedges are set to $t_\gamma = 0.5$, meaning that a hyperedge becomes activated when at least half of its nodes are already active. At the initial time $t=0$, node $3$ is chosen as the initial seed, so $U(0)=\{3\}$. Depending on the threshold rule, the hyperedges $e_{{\gamma}_2}$ and $e_{{\gamma}_3}$ get activated at time $t=1$ because their fractions of activated nodes reached threshold $0.5$, so $U(1)=\{3, e_{{\gamma}_2}, e_{{\gamma}_3}\}$. At time $t=2$, nodes $2$ and $6$ are made to activated subsequently and $U(2)=\{3, e_{{\gamma}_2},e_{{\gamma}_3}, 2, 6\}$. At next step, the hyperedges $e_{{\gamma}_1}$ and $e_{{\gamma}_5}$ are both activated, making nodes $1$ and $7$ activated respectively, so $U(3)=\{3, e_{{\gamma}_2},e_{{\gamma}_3}, 2, 6, e_{{\gamma}_1},e_{{\gamma}_5}\}$ and $U(4)=\{3, e_{{\gamma}_2},e_{{\gamma}_3}, 2, 6, e_{{\gamma}_1},e_{{\gamma}_5}, 1, 7\}$. Since no more nodes or hyperedges will be activated, this propagation process ends at time $t=4$. 

In this paper, all thresholds of the hyperedges are set to the same value, which range from $0.5$ to $0.8$. 

\section{Theoretical Analysis and HCI-TM Algorithm}

In this Section, we propose a collective influence theoretical analysis framework based on threshold models in hypergraphs.

\subsection{Message Passing Equation}

In the spread process, $v_i$ represents the probability of node $i$ being activated, and $v_{e_{\gamma}}$ the probability of hyperedge $e_{\gamma}$ being activated. At the final stage of propagation, $v_i$ and $v_{e_{\gamma}}$ can only show two states, that is, $v_i=0(v_{e_\gamma}=0)$, indicating that node $i$ (hyperedge $e_\gamma$) has not been activated, and $v_i=1(v_{e_\gamma}=1)$, indicating that node $i$ (hyperedge $e_\gamma$) has been activated. Therefore, we denote $Q(q) = \sum\limits_{i = 1}^N {\frac{{{v_i}}}{N}} $ as the final proportion of the active nodes with size $q$ of the initial seed nodes. 

In hypergraph, the Cavity Method can be extended to the "link" between node and hyperedge. Actually, the hypergraph can be described as a Bipartite Graph, which consist of two kinds of node: the nodes and hyperedges in hypergraph are both taken as the node in Bipartite Graph, the links in Bipartite Graph only exist between node and hyperedge when they are associated in hypergraph. So we can consider to remove two kinds of node in Bipartite Graph by Cavity Method. For a directed link from node $i$ to hyperedge $e_\gamma$, suppose $e_\gamma$ is "virtually" removed from the Bipartite Graph and reconsider if node $i$ is activated or not. The variable $v_{i \to {e_\gamma }}^{t}$ represents the probability of node $i$ being activated in the absence of hyperedge $e_\gamma$ at time $t$. Similarly, $v_{{e_\gamma } \to i}^{t}$ represents the probability of hyperedge $e_\gamma$ being activated in the absence of node $i$ at time $t$. Notice that although the direct impact between node $i$ and hyperedge $e_\gamma$ did not occur, they can still affect each other by a longer path $i \leftrightarrow {e_\beta } \leftrightarrow \ldots \leftrightarrow j \leftrightarrow {e_\gamma }$. For sparse hypergraphs, the updating process can be obtained:

\begin{align}
\left\{ \begin{array}{l}
{v_{i \to {e_\gamma }}^{t+1}} = {n_i} + (1 - {n_i})[1 - \prod\limits_{{e_{\beta}} \in \partial i/{e_\gamma }} {(1 - {v_{{e_{\beta}} \to i}^{t}})} ]\\
{v_{{e_\gamma } \to i}^{t+1}} = 1 - \prod\limits_{{P_h} \in P_{{e_\gamma }/i}^{{m_\gamma }}} {(1 - \prod\limits_{p \in {P_h}} {{v_{p \to {e_\gamma }}^{t}}} )} 
\end{array} \right.
\label{Cavity}
\end{align}

Here, $\partial i/{e_\gamma }$ represents the set of hyperedges related to node $i$ except $e_\gamma$. $P_{{e_\gamma }/i}^{{m_\gamma }} = \{ {P_1},{P_2},...,{P_\tau }\}$ is defined as the set of all combinations of $m_\gamma$ nodes in hyperedge $e_\gamma$ excluding node $i$, where $\tau  = C_{{N_\gamma } - 1}^{{m_\gamma }}$. Each element ${P_h}$ in $P_{{e_\gamma }/i}^{{m_\gamma }}$ has the form ${P_h} = \{ {p_{h_1}},{p_{h_2}},...,{p_{h_{m_\gamma }}}\}$, representing the $h$th combination composed of $m_\gamma$ nodes. Notice that in Eq.(\ref{Cavity}), the states of $v_{i\to e_{\gamma}}^{t}$ and $v_{e_{\gamma}\to i}^{t}$ will update every two time steps (i.e. $v_{i\to e_{\gamma}}^{2t+1}=v_{i\to e_{\gamma}}^{2t}$ and $v_{e_{\gamma}\to i}^{2t+2}=v_{e_{\gamma}\to i}^{2t+1}$ for all $t=0,1,\ldots$). Denote ${\lim_{t\to \infty} v_{i \to {e_\gamma }}^{t}=v_{i \to {e_\gamma }}}$ and ${\lim_{t\to \infty} v_{{e_\gamma } \to i}^{t}=v_{{e_\gamma } \to i}}$. The message passing equations can be described as:

\begin{align}
\left\{ \begin{array}{l}
{v_{i \to {e_\gamma }}} = {n_i} + (1 - {n_i})[1 - \prod\limits_{{e_\beta} \in \partial i/{e_\gamma }} {(1 - {v_{{e_\beta} \to i}})} ]\\
{v_{{e_\gamma } \to i}} = 1 - \prod\limits_{{P_h} \in P_{{e_\gamma }/i}^{{m_\gamma }}} {(1 - \prod\limits_{p \in {P_h}} {{v_{p \to {e_\gamma }}}} )} 
\end{array} \right.
\label{MP}
\end{align}

Moreover, the final state of node $i$ and hyperedge $e_\gamma$ is given as follows:

\begin{align}
\left\{ \begin{array}{l}
{v_i} = {n_i} + (1 - {n_i})[1 - \prod\limits_{{e_\beta} \in \partial i} {(1 - {v_{{e_\beta}\to i}})} ]\\
{v_{{e_\gamma }}} = 1 - \prod\limits_{{P_h} \in P_{{e_\gamma }}^{{m_\gamma }}} {(1 - \prod\limits_{p \in {P_h}} {{v_{p \to e_{\gamma}}}} )} 
\end{array} \right.
\label{state}
\end{align}

\subsection{High-order Collective Influence}

To simplify the Eq.(\ref{MP}), let's denote $\mathbf{V}_{\to} = \{\textbf{v}_{1},\textbf{v}_{2}\}^T$, where $\textbf{v}_{1}=\{v_{i \to {e_\gamma }}\}_{S \times 1}$, $\textbf{v}_{2}=\{v_{{e_\gamma } \to i}\}_{S \times 1}$. Here $S = \sum\limits_{i = 1}^N {{k_i}}$ and $k_i$ represents the hyperdegree of node $i$. Correspondingly, we extend $\textbf{n}$ to $\textbf{n}_{\to}$ with larger dimension $2S$: 

\begin{align}
\textbf{n}_{\to} = (\textbf{n}_1, \textbf{0})^{T} = (\underbrace {\ldots,\overbrace{{n_i},\ldots,{n_i}}^{k_i},\ldots}_S, \underbrace {0,\ldots,0}_S)^T
\label{n}
\end{align}

So the Eq.(\ref{MP}) can be described as a nonlinear function:

\begin{align}
\mathbf{V}_{\to} = \textbf{n}_{\to} + \mathcal{G}(\mathbf{V}_{\to}) \Leftrightarrow
\left\{ \begin{array}{llll}
\mathbf{v}_1 &=\textbf{n}_{1} &+ &\textbf{g}_1({\textbf{v}}_2)\\
\mathbf{v}_2 &=\textbf{0} &+ &\textbf{g}_2({\textbf{v}}_1)
\end{array} \right.
\label{Selfconsistent}
\end{align}

The final state of $\mathbf{V}_{\to}$ is completely determined by giving the initial seeds set $\textbf{n}_{\to}$, i.e. the solution of the self-consistent Eq.({\ref{Selfconsistent}}) is unique under given initial seeds. Unfortunately, it is hard to solved directly due to the complexity of functions $\textbf{g}_1$ and $\textbf{g}_2$\cite{2016Efficient}, Therefore, we calculate it approximately by iteration and linearization:

\begin{align}
\mathbf{V}_{\to}^{t+1} = \textbf{n}_{\to} + J\mathcal{G}|_{\mathbf{V}_{\to}^{t}}\times\mathbf{V}_{\to}^{t}
\label{V_iter}
\end{align}

Here the $J\mathcal{G}|_{\mathbf{V}_{\to}^{t}}$ is the jacobian matrix of Eq.(\ref{Selfconsistent}) at point $\mathbf{V}_{\to}^{t}$. For a given hypergraph, we take the partial derivatives of Eq.(\ref{Selfconsistent}):

\begin{align}
J\mathcal{G} = \left( \begin{array}{cc}
\frac{\partial \textbf{g}_1}{\partial  {\textbf{v}}_1}&\frac{\partial \textbf{g}_1}{\partial  {\textbf{v}}_2}\\[2mm]
\frac{\partial \textbf{g}_2}{\partial  {\textbf{v}}_1}&\frac{\partial \textbf{g}_2}{\partial  {\textbf{v}}_2}
\end{array} \right)_{2S \times 2S}
\label{Jacobian}
\end{align}

For the partial derivative of $\textbf{g}_1$, We have:

\begin{align}
\frac{\partial v_{i \to {e_\gamma}}}{\partial v_{j \to {e_\beta}}} &= 0 \label{g11} \\
\frac{\partial v_{i \to {e_\gamma}}}{\partial v_{e_\beta \to j}}  &=\left\{ \begin{array}{cl}
(1-n_i)\prod\limits_{e_\mu \in \partial i/e_\gamma,e_\beta} (1-v_{e_\mu \to i})  & i = j,{e_\beta } \ne {e_\gamma }\\
0  & \text{otherwise}
\end{array} \right. 
\label{g12}
\end{align}

For the first row of jacobian matrix at time $t$, we have $\frac{\partial \textbf{g}_1}{\partial {\textbf{v}}_1}\big\vert_{\mathbf{V}_{\to}^{t}}=\textbf{0}$ and $\frac{\partial \textbf{g}_1}{\partial {\textbf{v}}_2}\big\vert_{\mathbf{V}_{\to}^{t}} = \{\frac{\partial v_{i \to {e_\gamma}}}{\partial v_{e_\beta \to j}}\}\big\vert_{\mathbf{V}_{\to}^{t}}$ is a generalization of the non-backtracking (NB) matrix, which is only determined by the number of active hyperedges associating to node $i$ except $e_\gamma$ and $e_\beta$ at time $t$, denoted by $a^t_{e_\beta \to i,i \to e_\gamma}= \sum\limits_{e_\mu \in \partial {i }/(e_\gamma,e_\beta)} {{v_{e_\mu \to i}^{t}}}$. So the $\frac{\partial \textbf{g}_1}{\partial {\textbf{v}}_2}$ can be described as:

\begin{align}
\frac{\partial v_{i \to {e_\gamma}}}{\partial v_{e_\beta \to j}}\bigg\vert_{\mathbf{V}_{\to}^{t}} = \left\{ \begin{array}{cl}
1-n_{i}  &\quad i = j,{e_\beta } \ne {e_\gamma }, a_{e_\beta \to i,i \to e_\gamma}^{t}=0\\
0  &\quad\text{otherwise}
\end{array} \right.
\label{NB_M}
\end{align}

The same analysis on the partial derivative of $\textbf{g}_2$ yields that $\frac{{\partial {v_{{e_\gamma } \to i}}}}{{\partial {v_{{e_\beta } \to j}}}}$ is always zero and $\frac{{\partial {v_{{e_\gamma } \to i}}}}{{\partial {v_{j \to {e_\beta }}}}}$ is almost zero except for ${e_\beta } = {e_\gamma },j \ne i$:

\begin{align}
\frac{{\partial {v_{{e_\gamma } \to i}}}}{{\partial {v_{j \to {e_\gamma }}}}} &= \prod\limits_{\scriptstyle{P_h} \in P_{{e_\gamma }/i}^{{m_\gamma }}\hfill\atop
\scriptstyle j \notin {P_h}\hfill} {(1 - \prod\limits_{p \in {P_h}} {{v_{p \to {e_\gamma }}}} )}\notag\\  &\times \sum\limits_{\scriptstyle{P_h} \in P_{{e_\gamma }/i}^{{m_\gamma }}\hfill\atop
\scriptstyle j \in {P_h}\hfill} {[\prod\limits_{p \in {P_h}/j} {{v_{p \to {e_\gamma }}}} \prod\limits_{\scriptstyle{{\tilde P}_h} \in P_{{e_\gamma }/i}^{{m_\gamma }}\hfill\atop
{\scriptstyle{{\tilde P}_h} \ne {P_h}\hfill\atop
\scriptstyle j \in {{\tilde P}_h}\hfill}} {(1 - \prod\limits_{p \in {{\tilde P}_h}} {{v_{p \to {e_\gamma }}}} )} ]} \label{g21}
\end{align}

Although the $\frac{{\partial {v_{{e_\gamma } \to i}}}}{{\partial {v_{j \to {e_\gamma }}}}}\big\vert_{\mathbf{V}_{\to}^{t}}$ seems much complicated, it still can be simplified by the $b^t_{j \to {e_\gamma},{e_\gamma } \to i} = \sum\limits_{p \in \partial {e_\gamma }/(i,j)} {{v^t_{p \to {e_\gamma }}}} $, which is the number of active nodes associating to $e_\gamma$ at time $t$, excluding $i$ and $j$. Firstly, if $b^t_{j \to {e_\gamma},{e_\gamma } \to i}  \ge {m_\gamma }$ at time $t$, there is at least one combination of $P_h$ that makes for $\prod\limits_{p \in {P_h}} {{v^t_{p \to {e_\gamma }}}}  = 1$, which leads to that the first part on the right side of Eq.(\ref{g21}) equals zero, so $\frac{{\partial {v_{{e_\gamma } \to i}}}}{{\partial {v_{j \to {e_\gamma }}}}}\big\vert_{\mathbf{V}_{\to}^{t}}=0$ under this condition. Secondly, if $b^t_{j \to {e_\gamma},{e_\gamma } \to i} \le {m_\gamma } - 2$ at time $t$, we have $\prod\limits_{p \in {P_h}/j} {{v^t_{p \to {e_\gamma }}}} =0$ for any combination because at least one zero element will be selected when you choose $m_{\gamma}-1$ elements from a set containing at most $m_{\gamma}-2$ nonzero elements, so $\frac{{\partial {v_{{e_\gamma } \to i}}}}{{\partial {v_{j \to {e_\gamma }}}}}\big\vert_{\mathbf{V}_{\to}^{t}}=0$ under this condition. Lastly, only when $b^t_{j \to {e_\gamma},{e_\gamma } \to i} = {m_\gamma } - 1$, there is exactly one combination such that $\prod\limits_{p \in {P_h}/j} {{v^t_{p \to {e_\gamma }}}}  = 1$. Meanwhile, all the forms $\prod\limits_{p \in {P_h}} {{v^t_{p \to {e_\gamma }}}}$ and $\prod\limits_{p \in {{\tilde P}_h}} {{v^t_{p \to {e_\gamma }}}}$ are zeros, which finally leads to $\frac{{\partial {v_{{e_\gamma } \to i}}}}{{\partial {v_{j \to {e_\gamma }}}}}\big\vert_{\mathbf{V}_{\to}^{t}} =1$. Therefore, for the second row of jacobian matrix at time $t$, we have $\frac{\partial \textbf{g}_2}{\partial  {\textbf{v}}_2}\big\vert_{\mathbf{V}_{\to}^{t}}=\textbf{0}$ and another generalized Non-Backtracking matrix $\frac{\partial \textbf{g}_2}{\partial  {\textbf{v}}_1}\big\vert_{\mathbf{V}_{\to}^{t}} = \{\frac{{\partial {v_{{e_\gamma } \to i}}}}{{\partial {v_{j \to {e_\beta }}}}}\}\big\vert_{\mathbf{V}_{\to}^{t}}$ described as:

\begin{align}
\frac{{\partial {v_{{e_\gamma } \to i}}}}{{\partial {v_{j \to {e_\beta }}}}}\bigg\vert_{\mathbf{V}_{\to}^{t}}= \left\{ \begin{array}{ll}
1  &\quad {e_\beta } = {e_\gamma },j \ne i,{b^t_{j \to {e_\gamma},{e_\gamma } \to i}} = {m_\gamma } - 1\\
0  &\quad \text{otherwise}
\end{array} \right.
\label{NB_I}
\end{align}

For the convenience and simplicity of the following derivation, the jacobian matrix $J\mathcal{G}\big\vert_{\mathbf{V}_{\to}^{t}}$ is denoted as follows:

\begin{align}
J\mathcal{G}\big\vert_{\mathbf{V}_{\to}^{t}} = \left( {\begin{array}{*{20}{c}}
\textbf{0}&\mathcal{M}^{t}\\
\mathcal{I}^{t}&\textbf{0}
\end{array}} \right)
\end{align}

Here $\mathcal{M}^{t}=\{\mathcal{M}^t_{e_\beta \to j,i \to e_\gamma}\}=\{\frac{\partial v_{i \to {e_\gamma}}}{\partial v_{e_\beta \to j}}\}\bigg\vert_{\mathbf{V}_{\to}^{t}}$, $\mathcal{I}^{t}=\{\mathcal{I}_{j \to e_\beta,e_\gamma \to i}\}\big\vert_{\mathbf{V}_{\to}^{t}}=\{\frac{\partial v_{e_\gamma \to i}}{\partial v_{j \to e_\beta}}\}\bigg\vert_{\mathbf{V}_{\to}^{t}}$ are the matrix with demension $S\times S$. We extend the matrices $\mathcal{M}^t$ and $\mathcal{I}^t$ to the higher dimensional space $M\times N \times N \times M$ and $N\times M \times M \times N$. Although they should not be combined to form a matrix because of their different dimensions, it does not affect the subsequent calculation as long as we just do it respectively. Bringing in new symbols:

\begin{align}
\left\{ \begin{array}{l}
\mathcal{M}^t_{e_{\beta} ji e_{\gamma}} = (1 - {n_i}){H_{i{e_\gamma }}}{H_{j{e_\beta }}}{\delta _{ij}}(1 - {\delta _{{e_\beta }{e_\gamma }}})M^t_{{e_\beta }ii{e_\gamma }}\\
{\mathcal{I}^t_{j{e_\beta }{e_\gamma }i}} = {H_{i{e_\gamma }}}{H_{j{e_\beta }}}{\delta _{{e_\beta }{e_\gamma }}}(1 - {\delta _{ij}}){I^t_{j{e_\gamma }{e_\gamma }i}}
\end{array} \right.
\end{align}

Here $M^t_{{e_\beta }ii{e_\gamma }}$ and ${I^t_{j{e_\gamma }{e_\gamma }i}}$ are the binary matrices. $M^t_{{e_\beta }ii{e_\gamma }}=1$ if $a^t_{e_\beta \to i,i \to e_\gamma}=0$ and $M^t_{{e_\beta }ii{e_\gamma }}=0$ otherwise. Similarly, ${I^t_{j{e_\gamma }{e_\gamma }i}}=1$ if $b^t_{j \to {e_\gamma},{e_\gamma } \to i} =m_{\gamma}-1$ and ${I^t_{j{e_\gamma }{e_\gamma }i}}=0$ otherwise. The $M^t_{{e_\beta }ii{e_\gamma }}$ and ${I^t_{j{e_\gamma }{e_\gamma }i}}$ are related to the concept of "subcritical" node and hyperedge respectively. By definition of Ref\cite{2016Efficient}, a node (hyperedge) is subcritical if that one more activation of associated hyperedge (node) will cause it activated. Additionally, $M^t_{{e_\beta }ii{e_\gamma }}$ and ${I^t_{j{e_\gamma }{e_\gamma }i}}$ are fully determined by $v^t_{i \to e_\gamma}$ and $v^t_{e_\gamma \to i}$, implying that both of them are also periodic (${M}^{2t+2}_{{e_\beta }ii{e_\gamma }}={M}^{2t+1}_{{e_\beta }ii{e_\gamma }}$ and $I_{j{e_\gamma }{e_\gamma }i}^{2t+1}=I_{j{e_\gamma }{e_\gamma }i}^{2t}$ for all $t=0,1,\ldots$).

For $t=1$, we set $\mathbf{V}_{\to}^{0} = \textbf{n}_{\to}$, so $\mathbf{V}_{\to}^{1}= \textbf{n}_{\to}+ J\mathcal{G}^{0}\times\textbf{n}_{\to}$:

\begin{align}
{\left[ {\begin{array}{*{20}{c}}
\textbf{v}_{1}\\
\textbf{v}_{2}
\end{array}} \right]^1} = \left[ {\begin{array}{*{20}{c}}
\textbf{n}_{1}\\
\textbf{0}
\end{array}} \right] + \left[ {\begin{array}{*{20}{c}}
0& \mathcal{M}^{0}\\
\mathcal{I}^{0}&0
\end{array}} \right]\left[ {\begin{array}{*{20}{c}}
\textbf{n}_{1}\\
\textbf{0}
\end{array}} \right] = \left[ {\begin{array}{*{20}{c}}
\textbf{n}_{1}\\
{\mathcal{I}^0}\textbf{n}_{1}
\end{array}} \right]
\label{MP1}
\end{align}

Notice that $\textbf{n}_{1}$ should also been extended to the higher dimensional space $N \times M$, based on the Eq.(\ref{MP1}), we have: 

\begin{figure*}
	\centering
		\includegraphics[scale=.2]{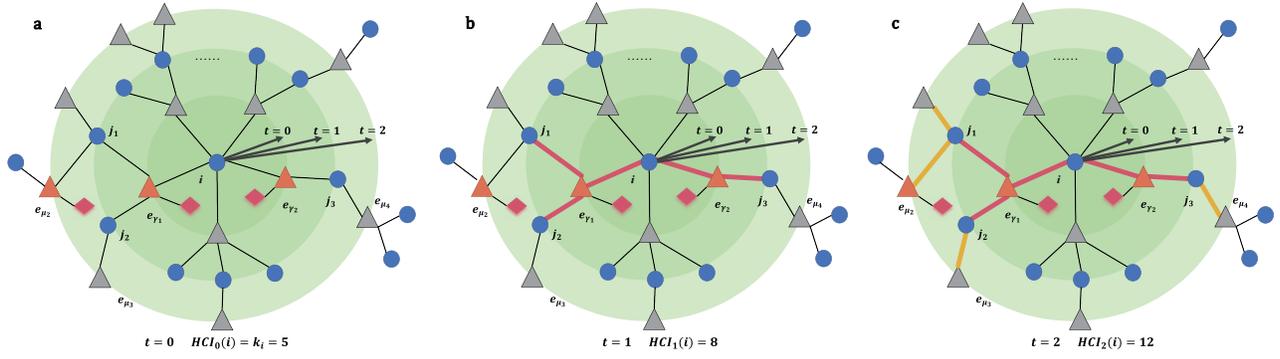}
	\caption{The illustrates of subcritical path in hypergraphs. The blue circles represent nodes, the triangles represent hyperedges. (a) At t=0, HC$\rm I_{0}(i)$ is erqual to the hyperdegree of node i. (b) At t=1, the additional 2-length subcritical paths also contribute to HC$\rm I_{1}(i)$, which represented by thick crimson links. And the orange triangles represent hyperedges in subcritical state, so the HC$\rm I_{1}(i)$=HC$\rm I_{0}(i)$+3=8. (c) At t=2, the additional 3-length subcritical paths are represented by thick yellow line, so HC$\rm I_{2}(i)$=HC$\rm I_{1}(i)$+4=5+3+4=12.}
	\label{FIG:2}
\end{figure*}

\begin{align}
\left\{ \begin{array}{l}
v^1_{i \to {e_\gamma }} = n_i H_{i e_\gamma}\\
v^1_{{e_\gamma } \to i} = {H_{i{e_\gamma }}}\sum\limits_j {{n_j}{H_{j{e_\gamma }}}(1 - {\delta _{ij}})I_{j{e_\gamma }{e_\gamma }i}^0} 
\end{array} \right.
\label{V1}
\end{align}

Defining $\left\| {{v_ \to }} \right\| = \sum\limits_{i{e_\gamma }} ({v_{i \to {e_\gamma }}}+{v_{{e_\gamma } \to i}})$ to quantify the activation scale of Threshold Models in hypergraphs, our goal is to optimize the collective influence of a given number of seeds by maximizing $\left\| {{v_ \to }} \right\|$. Based on the Eq.(\ref{V1}), the form of $\left\| {{v_ \to }} \right\|$ for $t=1$ is:

\begin{align}
\left\| {{v_ \to }} \right\| &= \sum\limits_{i{e_\gamma }} {{v_{i \to {e_\gamma }}}}  + \sum\limits_{i{e_\gamma }} {{v_{{e_\gamma } \to i}}}\notag \\
&= \sum\limits_{i{e_\gamma }} {{n_i}H_{i{e_\gamma }}}  + \sum\limits_{i{e_\gamma }} {{H_{i{e_\gamma }}}} \sum\limits_j {n_j{H_{j{e_\gamma }}}(1 - {\delta _{ij}})I_{j{e_\gamma }{e_\gamma }i}^0}\notag\\
&= \sum\limits_i {{n_i}{k_i}}  + \sum\limits_i {{n_i}} \sum\limits_{{e_\gamma } \in \partial i} {\sum\limits_{j \in \partial{e_\gamma }/i} {I_{i{e_\gamma }{e_\gamma }j}^0} }\notag\\
&= \sum\limits_i {{n_i}} \left({k_i} + \sum\limits_{{e_\gamma } \in \partial i} {\sum\limits_{j \in \partial{e_\gamma }/i} {I_{i{e_\gamma }{e_\gamma }j}^0} } \right)
\label{V_sum_1}
\end{align}

Supposing that the final state of self-consistent Eq.(\ref{Selfconsistent}) is obtained at $t=1$, we can quantify the contribution of node $i$ to $\left\| {{v_ \to }} \right\|$ for $t=1$, which is formulated by the content in brackets of Eq.(\ref{V_sum_1}). Let's define it as the Hypergraph Collective Influence (HCI) of node $i$ to find the optimal influencers:

\begin{equation}
\begin{split}
HC{I_1}(i) &=  {k_i} + \sum\limits_{{e_\gamma } \in \partial i} {\sum\limits_{j \in \partial{e_\gamma }/i} {I_{i{e_\gamma }{e_\gamma }j}^0} } \\
\label{HCI1}
\end{split}
\end{equation}

As shown in Fig.(\ref{FIG:2}b), we can calculate the HC$\rm I_1 (i)$ through Eq.(\ref{HCI1}), so HC$\rm{I_1}(i)=8$. 

For $t=2$, based on Eq.(\ref{V_iter}), we have $\mathbf{V}_{\to}^{2}= \textbf{n}_{\to}+ J\mathcal{G}^{1}\times\mathbf{V}_{\to}^{1}=\mathbf{n}_{\to} + J\mathcal{G}^{1}\times \mathbf{n}_{\to} + J\mathcal{G}^{1}\times J\mathcal{G}^{0}\times \mathbf{n}_{\to}$, so:   

\begin{align}
\begin{split}
{\left[ {\begin{array}{*{20}{c}}
{\textbf {v}_1}\\
{\textbf {v}_2}
\end{array}} \right]^2}
= \left[ {\begin{array}{*{20}{c}}
{{\textbf {n}_1} + \mathcal{M}{^1{\mathcal{I}^0}}{\textbf {n}_1}}\\
{\mathcal{I}^1{\textbf { n}_1}}
\end{array}} \right]
\end{split}
\label{MP2}
\end{align}

Same as above, the specific form of Eq.(\ref{MP2}) can be described as

\begin{align}
\left\{ \begin{array}{lll}
v_{i \to {e_\gamma }}^{2} &=& n_{i}H_{i e_\gamma}+(1 - {n_i}){H_{i{e_\gamma }}}\sum\limits_{{e_\beta }} {{H_{i{e_\beta }}}(1-\delta_{e_\gamma e_\beta})M_{{e_\beta }ii{e_\gamma }}^1}\\
& &{\times \sum\limits_k {{n_k}{H_{k{e_\beta }}}(1 - {\delta _{ki}})I_{k{e_\beta }{e_\beta }i}^0} } \\
v_{{e_\gamma } \to i}^{2} &=& {H_{i{e_\gamma }}}\sum\limits_j {{n_j}{H_{j{e_\gamma }}}(1 - {\delta _{ij}})I_{j{e_\gamma }{e_\gamma }i}^1}
\end{array} \right.
\end{align}

Let us compute the Hypergraph Collective Influence of node $i$ for $t=2$, the same process as above, we have:

\begin{equation}
\begin{split}
HC{I_2}(i) &= {k_i} + \sum\limits_{{e_\gamma } \in \partial i} {\sum\limits_{j \in \partial{e_\gamma }/i} {I_{i{e_\gamma }{e_\gamma }j}^1} }  \\
&+ \sum\limits_{{e_\gamma } \in \partial i} {\sum\limits_{j \in \partial{e_\gamma }/i}I_{i{e_\gamma }{e_\gamma }j}^0 {\sum\limits_{{e_\mu } \in \partial j/{e_\gamma }} (1 - {n_j}) M_{{e_\gamma }jj{e_\mu }}^1} }   \\
\end{split}
\label{HCI2}
\end{equation}

As shown in Fig.(\ref{FIG:2}c), the additional 3-length paths also contribute to HC$\rm 
 I_2$(i), making HC$\rm I_2$(i)=12. Due to the periodicity of $I_{i{e_\gamma }{e_\gamma }j}$, we can replace the $I_{i{e_\gamma }{e_\gamma }j}^1$ with $I_{i{e_\gamma }{e_\gamma }j}^0$. Therefore, the right part of the first line in Eq.(\ref{HCI2}) is actually the HC$\rm I_{1}$(i). Moreover, $k_i$ can also be seen as HC$\rm I_{0}$(i), which corresponds to the High Hyper Degree (HHD) ranking. It is obvious that HC$\rm I_{t+1}$ is composed of HC$\rm I_{t}$ and all message of subcritical path with lengh $t$. So we can generalize the above HCI formula to any given $t=n$. In summary, based on Eq.(\ref{V_iter}), we have $\mathbf{V}_{\to}^{n} = \mathbf{n}_{\to} + {J\mathcal{G}^{n - 1}}\times \mathbf{V}_{\to}^{n - 1} = [1 + \sum\limits_{i = 1}^{n} \prod\limits_{j = 1}^{i} {J\mathcal{G}^{n - j}} ]\times \mathbf{n}_{\to}$, and for $n > 2$:

\begin{equation}
\begin{split}
{\left[ {\begin{array}{*{20}{c}}
{{\mathbf{v}_1}}\\
{{\mathbf{v}_2}}
\end{array}} \right]^n} = \left[ {\begin{array}{*{20}{c}}
{{\mathbf{n}_1} +\left( \sum\limits_{L \in A_n} {\prod\limits_{l \in {A_L}} {{\mathcal{M}^{n - l + 1}}{\mathcal{I}^{n - l}} } } \right){\mathbf{n}_1}}\\
{{\mathcal{I}^{n - 1}}\left({\mathbf{n}_1} + \left( \sum\limits_{L \in A_n} {\prod\limits_{l \in {A_L}} {{\mathcal{M}^{n - l}}{\mathcal{I}^{n - l - 1}}}} \right){\mathbf{n}_1}\right)}
\end{array}} \right]
\end{split}
\label{MPn}
\end{equation}

where $A_n=\{x\in N^+| x \; mod\; 2=0,x\le n\}$. By analyzing the Eq.(\ref{MPn}) we can get the Hypergraph Collective Influence of node $i$ for $t=n$:

\begin{equation}
\begin{split}
HC{I_n}(i) = k_i + \sum\limits_{L\in A_n} \mathbb{O}^{n}_{L} + \sum\limits_{L\in B_n}\mathbb{E}^{n}_{L}
\end{split}
\end{equation}

where $B_n=\{x\in N^+| x \; mod\; 2 = 1,x\le n\}$, $\mathbb{O}^{n}_{L}$ and $\mathbb{E}^{n}_{L}$ are defined as follows:

\begin{equation}
\begin{array}{l}
\mathbb{O}^{n}_{L} = \sum\limits_{e_{\gamma_1} \in\partial i_1} \sum\limits_{i_2 \in\partial e_{\gamma_1}/i_1} I^{n-L}_{i_1 e_{\gamma_1} e_{\gamma_1} i_2} \sum\limits_{e_{\gamma_2} \in\partial i_2 /e_{\gamma_1}} (1-n_{i_2})M^{n-L+1}_{e_{\gamma_1} i_2 i_2 e_{\gamma_2}}\\
\times \cdots \times \sum\limits_{i_\ell \in\partial e_{\gamma_{\ell-1}}/i_{\ell-1}} I^{n-2}_{i_{\ell-1} e_{\gamma_{\ell-1}} e_{\gamma_{\ell-1}} i_\ell} \sum\limits_{e_{\gamma_\ell} \in\partial i_\ell /e_{\gamma_{\ell-1}}} (1-n_{i_\ell})M^{n-1}_{e_{\gamma_{\ell-1}} i_\ell i_\ell e_{\gamma_\ell}}
\end{array}
\end{equation}

\begin{equation}
\begin{array}{ll}
\mathbb{E}^{n}_{L} = &\sum\limits_{e_{\gamma_1} \in\partial i_1} \sum\limits_{i_2 \in\partial e_{\gamma_1}/i_1} I^{n-L}_{i_1 e_{\gamma_1} e_{\gamma_1} i_2} \sum\limits_{e_{\gamma_2} \in\partial i_2 /e_{\gamma_1}} (1-n_{i_2})M^{n-L+1}_{e_{\gamma_1} i_2 i_2 e_{\gamma_2}}\\
&\times \cdots \times \sum\limits_{i_{\iota+1} \in\partial e_{\gamma_{\iota}}/i_{\iota}} I^{n-1}_{i_{\iota} e_{\gamma_{\iota}} e_{\gamma_{\iota}} i_{\iota+1} }
\end{array}
\end{equation}

Here $\ell=\frac{L+2}{2}$ and $\iota=\frac{L+1}{2}$. $\mathbb{O}^{n}_{L}$ and $\mathbb{E}^{n}_{L}$ represent the number of subcritical path starting from node $i$ with odd and even length respectively, and both length of them do not exceed $n+1$. Inspired by above, we define the concept of subcritical paths. For a directed link $i_1 \to e_{\gamma_{1}}  \to...\to e_{\gamma_{l}}$ is a subcritical path of length $2l-1$, if $n_{i_1}=1, n_{i_2}=0,...,n_{i_l}=0$,$I_{i_1 e_{\gamma_1}e_{\gamma_1} i_2}=1,M_{e_{\gamma_1} i_2 i_2 e_{\gamma_2}}=1,...,I_{i_{1-1} e_{\gamma_{l-1}} e_{\gamma_{l-1}} i_l}=1,M_{e_{\gamma_{l-1}} i_l i_l e_{\gamma_l}}=1$. Same as above, the directed link $i_1 \to e_{\gamma_{1}}  \to...\to i_{l}$ is a subcritical path of length $2l-2$, if  $n_{i_1}=1, n_{i_2}=0,...,n_{i_l}=0$,$I_{i_1 e_{\gamma_1}e_{\gamma_1} i_2}=1,M_{e_{\gamma_1} i_2 i_2 e_{\gamma_2}}=1,...,I_{i_{l-1} e_{\gamma_{l-1}} e_{\gamma_{l-1}} i_l}=1$. For locally tree-like hypergraphs, HC$\rm I_n$ can be approximately defined as the number of subcritical paths starting from i with length $0 \le l \le n$.

\subsection{HCI-TM Algorithm}

Our objective is to identify the most effective configuration of seeds $\mathbf{n}$, with a given size of $qN$, in order to maximize $\left\| {{v_ \to }} \right\|$. By increasing the number of iterations or the time parameter $n$, we can obtain increasingly accurate approximations of the final state defined by Eq.(\ref{Selfconsistent}). This signifies improved performance of HC$\rm I_n (i)$. However, it also leads to an increase in the complexity of HC$\rm I_n (i)$, as it becomes dependent on time and other seeds. In order to select the optimal influence node, we have devised an adaptive HCI-TM algorithm that follows a greedy approach based on the aforementioned analysis:

\begin{algorithm}
    \SetAlgoLined 
	\caption{HC$\rm I_n$-TM algorithm}
	\KwIn{ \\Hypergraph: $\mathbb{H} (V,E)$,\\Activation Radio: $a_r$}
	\KwOut{\\Seed Set: S}
    \textbf{Initialization:}
	S=$\emptyset $, 
    Calculate the HC$\rm I_n$ values of all nodes in the hypergraph\;
	\While {Q(q)<$a_r$}{
       \quad Select node $i$ with the highest HC$\rm I_n$ as the seed\;
       \quad $S = \{ i\}  \cup S$\;
       \quad Remove the newly activated node and hyperedge from the \\ \quad hypergraph\;
       \quad Recalculate the HC$\rm I_n$ value of $\left\lceil n/2 \right\rceil $-layer neighbor of \\ \quad node removed in the prevous step\;
	}
	\textbf{return} S
 \label{HCI-TM1}
\end{algorithm}
It is important to note that $a_r$ represents the expected proportion of active nodes. In order to save computational resources, it is not necessary to update the HCI for all nodes. Instead, we only need to recalculate the neighbors within the $\left\lceil n/2 \right\rceil$-layer of nodes that were activated in the previous step. This approach allows us to focus on the relevant nodes and optimize the efficiency of the algorithm.

\section{\label{sec:level1}Numerical simulation:}

In this section, we have conducted a comprehensive evaluation of the HCI-TM algorithm by comparing it with several classical algorithms, namely HHD (High Hyper Degree Algorithm), HHDA (High Hyper Degree Adaptive Algorithm), NP (Neighbor Preference Algorithm), NPA (Neighbor Preference Adaptive Algorithm), PageRank, and RA (Random Algorithm). To ensure the robustness of our findings, we performed simulations on both synthetic and real-world hypergraphs. Our results revealed that the HCI-TM algorithm outperformed the other algorithms in various aspects. Specifically, it achieved maximum activation of the hypergraph while selecting a minimal seed set, thereby demonstrating its superior performance in the threshold models. It is worth noting that the parameter $a_r$ was set to $0.9$ in all experiments. This choice takes into account the presence of challenging-to-activate peripheral nodes in the hypergraphs. 

\subsection{\label{sec:level2}Synthetic hypergraphs}

To evaluate the effectiveness and efficiency of the HCI-TM algorithm, a series of experiments were conducted on synthetic hypergraphs. These hypergraphs consisted of Erds-Rényi (ER) hypergraphs\cite{2016Measuring} with average hyperdegrees of 2 and 3, Scale-Free (SF) hypergraphs\cite{2023HDA} with power-law exponent of 1.5 and 2, as well as K-uniform hypergraphs with hyperedge sizes of 4 and 5. The experiments covered a range of hypergraph size, including 5000, 10000, 20000, 30000, 50000, and 100000.  The result of the experiment was averaged ten times to ensure the reliability of the results. All numerical simulations was conducted in giant connected component of hypergraphs based on our threshold rules.

\begin{figure*}
	\centering
		\includegraphics[scale=.43]{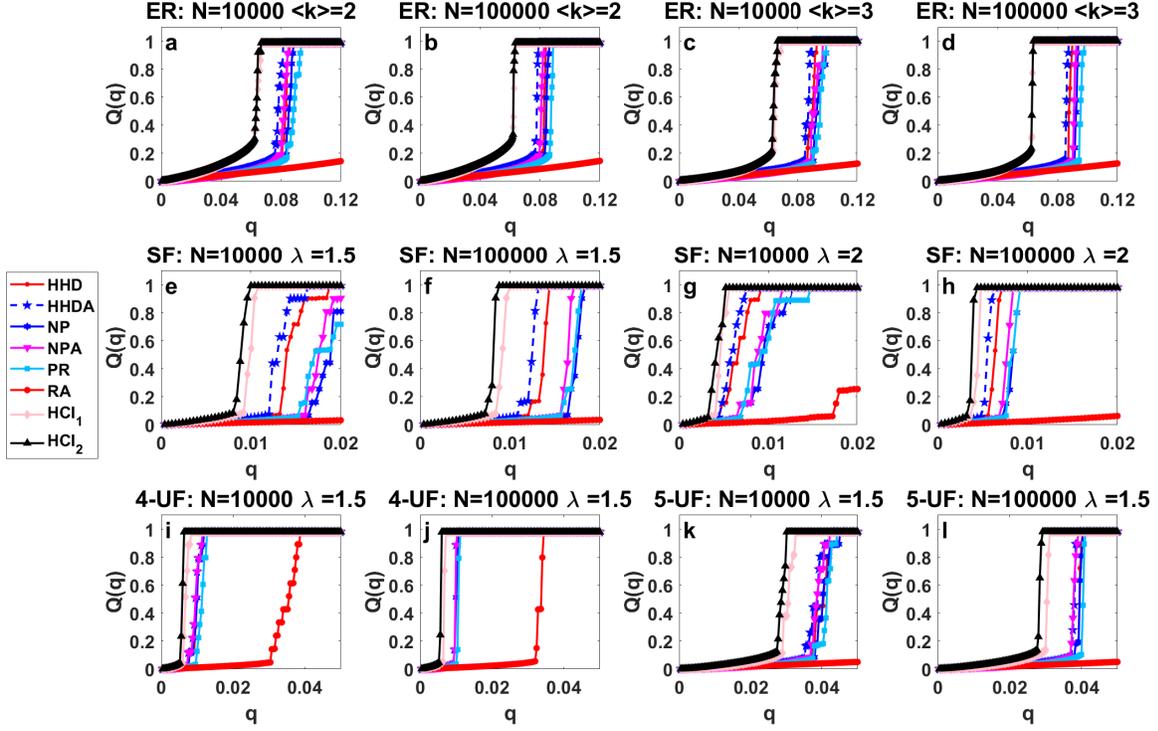}
	\caption{Performance of HCI-TM algorithm and other algorithms on ER, SF and K-UF hypergraphs. The size of hypergraphs N are 10000 and 100000, and $M=0.3N$ in ER hypergraphs, $M=0.5N$ in SF and K-uniform hypergraphs. The horizontal axi of each subfigure represents the proportion of seed nodes in the hypergraphs, and the vertical axis represents the proportion of active nodes in the hypergraphs.}
	\label{FIG:3}
\end{figure*}

FIG (\ref{FIG:3}) illustrates the performance of various algorithms in ER, SF, and K-UF hypergraphs under different parameter settings, with a threshold of $t_\gamma=0.5$. As depicted in FIG (\ref{FIG:3}), the HCI-TM algorithm outperforms other algorithms in all cases. It achieves a node activation rate of $90\%$ in the hypergraphs with the minimum number of seeds. The HHDA algorithm performs relatively good performance, ranking second in terms of effectiveness, while the RA algorithm need select the most seed nodes. Through our experimental simulations, we observed that the superiority of HCI-TM algorithm becomes increasingly evident in ER hypergraphs as the average hyperdegree increases. However, the performance of the HCI-TM algorithm on SF hypergraphs deteriorates as the power-law exponent increases. This phenomenon can be attributed to that the larger power-law exponent limits the number of nodes with large hyperdegrees, causing the more hyperdegree of nodes to be $1$, making the failure of the HCI-TM algorithm.

\begin{figure*}
	\centering
		\includegraphics[scale=.35]{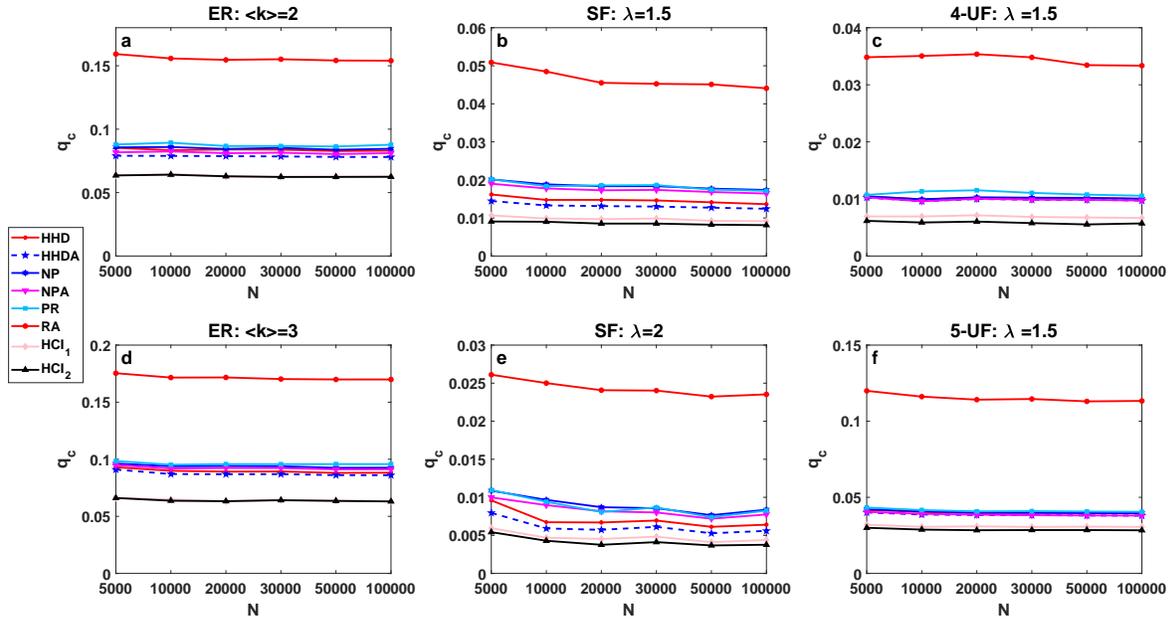}
	\caption{The fraction of seed nodes required by different algorithms to activate the ER hypergraphs with the average hyperdegree of 2 and 3, the SF hypergraphs with the power-law exponent of 1.5 and 2 and the K-uniform hypergraphs with the size of hyperedges are 4 and 5. The horizontal axis represents the hypergraph size, and the vertical axis represents the proportion of seed nodes.}
	\label{FIG:4}
\end{figure*}

FIG (\ref{FIG:4}) presents the proportion of seed nodes required by different algorithms to achieves 90\% node activation in ER, SF, and K-uniform hypergraphs. As shown in FIG (\ref{FIG:4}), the HCI-TM algorithm consistently outperforms other algorithms by selecting optimal influence seed sets with the minimum number of nodes. Following the HCI-TM algorithm, the HHDA algorithm demonstrates relatively good performance, while the RA algorithm requires the largest number of seed nodes.
These results highlight the effectiveness of the HCI-TM algorithm in identifying influential nodes in the hypergraphs, allowing it to achieve a high activation rate while minimizing resource utilization.

\begin{figure*}
	\centering
		\includegraphics[scale=.35]{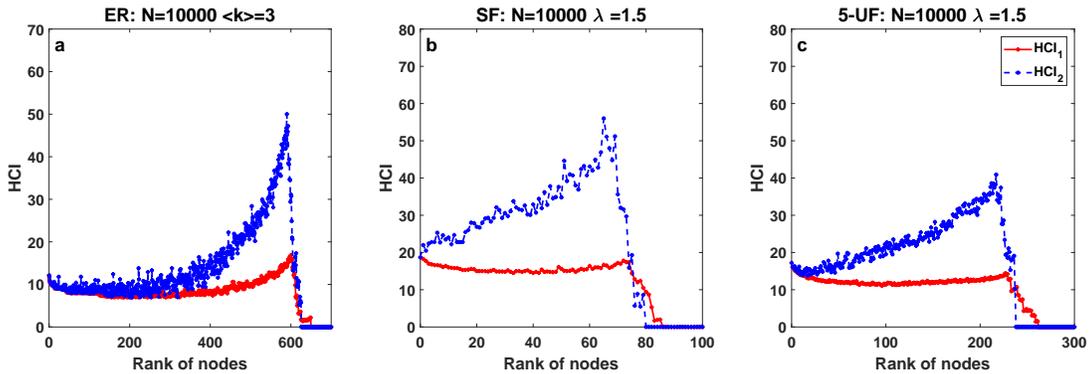}
	\caption{Analysis of the HCI value attached to each node when it is activated sequentially according to HC$\rm I_n$-TM ranking. a-c shows the results of HCI algorithm on ER hypergraphs with average hyperdegree of 3, SF hypergraphs with the power-law exponent of 1.5 and UF hypergraphs with size of hyperedges are 5, and the size of hypergraphs $N$ are $10000$.}
	\label{Fig:subpath}
\end{figure*}

In order to evaluate the effectiveness of HCI, we conducted an analysis of HC$\rm I_n$ during the spreading process. Initially, we examined the evolution of HCI based on the HCI-TM ranking during sequential activation, as shown in FIG (\ref{Fig:subpath}). It can be observed that in the early stages, the HCI value is small and there is a little disparity between HC$\rm I_1$ and HC$\rm I_2$. However, as the propagation process unfolds, the gap between HC$\rm I_1$ and HC$\rm I_2$ gradually widens. Furthermore, our findings indicate that the occurrence of the cascade phenomenon corresponds to the peak evolution of HCI. Notably, the peak of HC$\rm I_2$ occurs earlier and reaches a higher value compared to HC$\rm I_1$. Remarkably, the HC$\rm I_2$ algorithm achieves this superior performance while utilizing fewer seed nodes. These results demonstrate the significant impact of HCI in the cascading process and suggest that it can be used as a tool to predict the occurrence of cascading phenomen.

\begin{figure*}
	\centering
		\includegraphics[scale=.35]{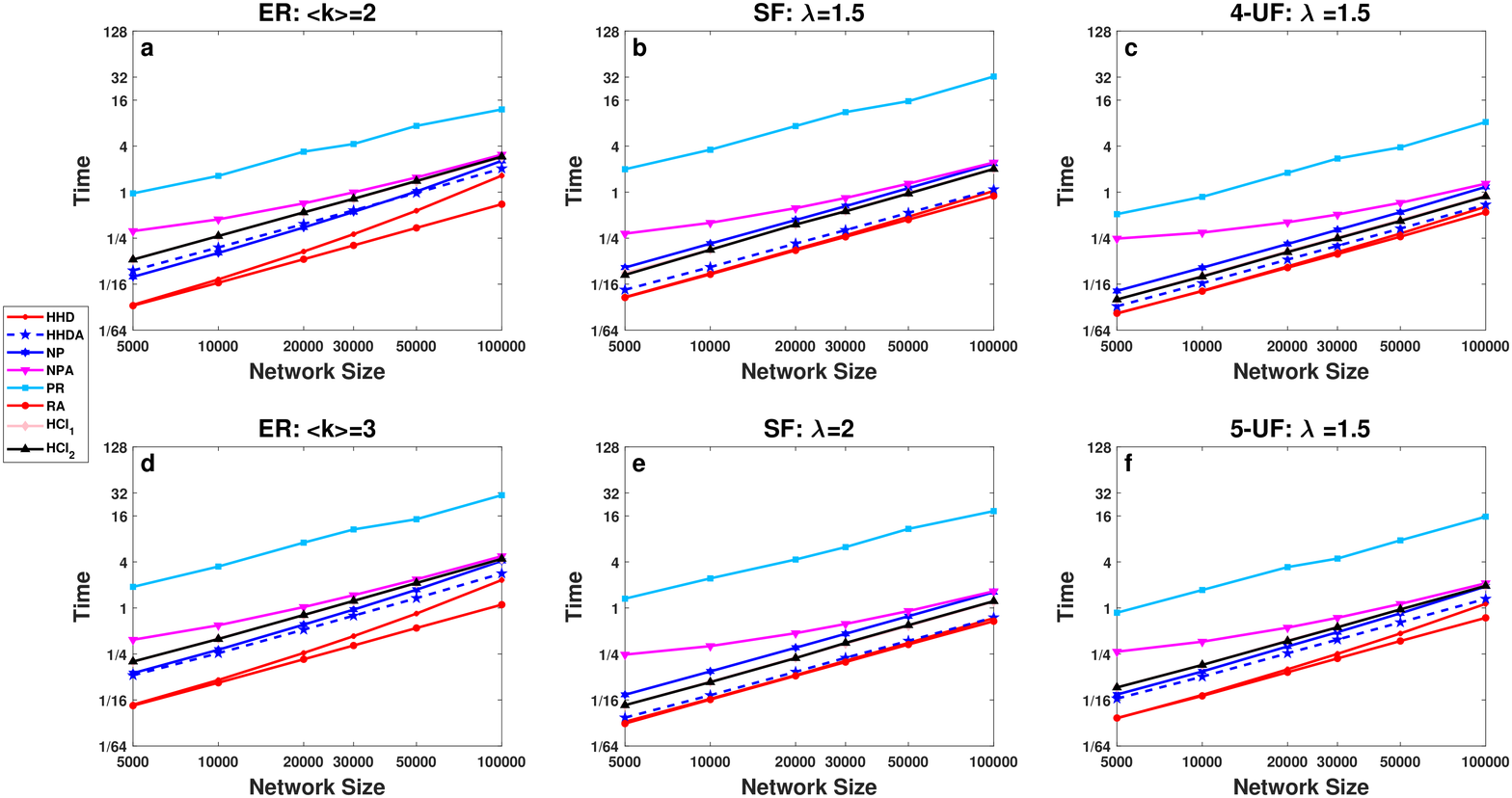}
	\caption{The computational time (seconds) of different algorithms on ER hypergraphs, SF hypergraphs and the K-uniform hypergraphs. The horizontal axis is $log_{10}(N)$, and the vertical axis is $log_{10}(T)$, where $N$ represent the number of node in the hypergraphs and $T$ represent computational time. }
	\label{FIG:5}
\end{figure*}

\begin{table}
\centering
\caption{The TABLE show the slope of different algorithm running times using a one order polynomial fitting on different types of hypergraphs.}
\label{tab:1}
\begin{tabular}{ccccccccc}
\hline
Algorithms& HHD& HHDA& NP &NPA &PR &RA &HC$\rm I_1$ &HC$\rm I_2$\\
\hline
ER-2& 1.28& 1.03& 1.17& 0.76& 0.87& 1.04& 1.02& 1.03\\
ER-3& 1.25& 1.03& 1.13& 0.83& 0.92& 1.02& 1.02& 1.03\\
SF-1.5& 1.07& 1.02& 1.06& 0.73& 0.95& 1.04& 1.07& 1.07\\
SF-2& 1.04& 1.01& 1.03& 0.64& 0.89& 1.02& 1.03& 1.06\\
4-UF& 1.06& 1.01& 1.04& 0.58& 0.92& 1.00& 1.02& 1.01\\
5-UF& 1.16& 1.02& 1.10& 0.70& 0.97& 1.04& 1.03& 1.04\\
\hline
\end{tabular}
\label{TC}
\end{table}

We recorded the computational time of each algorithm in different synthetic hypergraphs. The logarithm ($\log_{10}$) of the computational time and the scale of the hypergraph were then calculated. Subsequently, we plotted a line graph with the scale of hypergraph on the horizontal axis and the time on the vertical axis, as shown in FIG (\ref{FIG:5}). To analyze the time complexity of the algorithms, we fitted the running time data with first-order polynomials, as presented in TABLE (\ref{TC}). Our analysis revealed that the time consumed by each algorithm exhibits a linear growth pattern in relation to the hypergraph size. Remarkably, the HHD algorithm exhibits the highest time complexity, approximately $O(N^{1.14})$, while the NPA algorithm demonstrates the lowest time complexity, approximately $O(N^{0.71})$. The time complexity of the HCI-TM algorithm, approximately $O(N^{1.04})$, demonstrates a linear increase in computational time as the scale of the hypergraph increases. Among all the algorithms examined, the PageRank algorithm consumes the most time, while the RA algorithm exhibits the lowest time complexity. The HCI-TM algorithm strikes a reasonable balance between time complexity and performance, making it a pragmatic choice for practical applications.

\subsection{\label{sec:level2}Real-world hypergraphs}
The threshold models in hypergraphs is widely utilized in various real-world scenarios, encompassing the proliferation of public opinion and protein interaction hypergraphs. In this subsection, we conducted six numerical simulations in real-world hypergraphs to demonstrate the efficacy of the HCI-TM algorithm. The specific information about the real-world hypergraphs is presented in TABLE (\ref{TC2}).

\begin{table}[!h]
\centering
\caption{Some specific information about the real-world hypergrphs dataset, including the number of nodes and hyperedges, and the threshold settings.}
\label{tab:2}
\begin{tabular}{cccc}
\hline
Dataset& \quad N \quad& \quad M\quad& \quad $t_\gamma$ \quad \\
\hline
House-committees& 1,290& 341& 0.6\\
Vegas-bars-reviews& 1,234& 1,194& 0.6\\
Geometry-questions& 580& 1,193& 0.8\\
Senate-committees& 282& 315& 0.8\\
MAG-10& 80,198& 51,889& 0.5\\
Mathoverflow-answers& 73,851& 5,446& 0.5\\
\hline
\end{tabular}
\label{TC2}
\end{table}

The first dataset we used is Congressional data compiled by Charles Stewart and Jonathan Vaughan\cite{abh1303}. The nodes of the House-Committees dataset represent members of the U.S. house of Representatives, and the hyperedges represent the committee members. There are 1290 nodes and 341 hyperedges in the hypergraph, and the average and median number of nodes in each hyperedges are 34.8 and 40 respectively.

\begin{figure*}
	\centering
		\includegraphics[scale=.35]{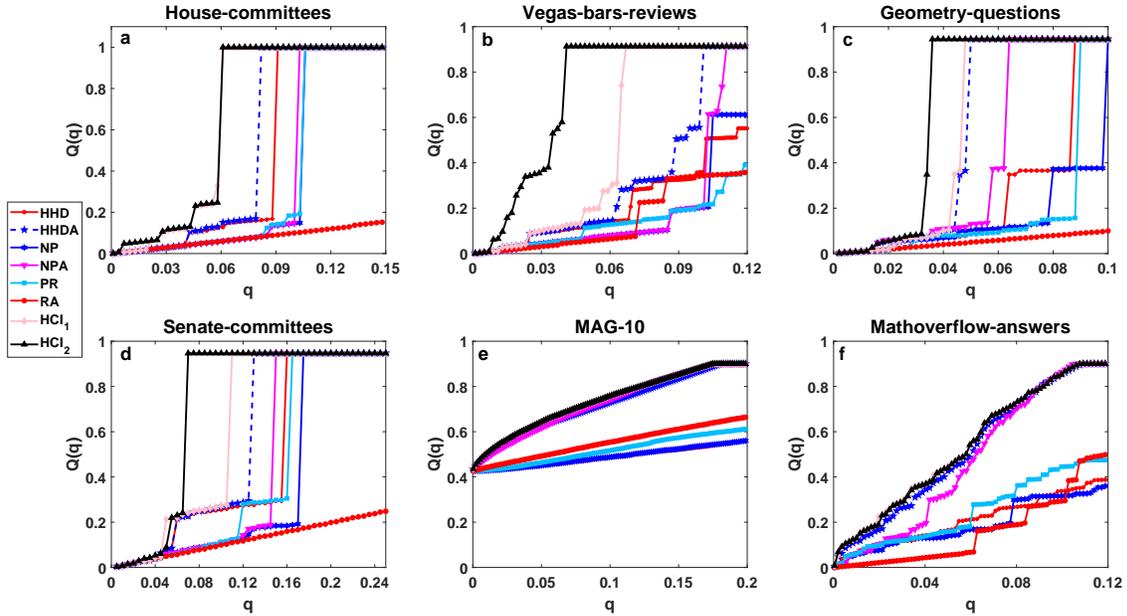}
	\caption{Performance of HCI-TM algorithm and other classical algorithms on real-world hypergraphs. And the horizontal axis represent the proportion of seed nodes and the vertical axis represent the proportion of active nodes in the hypergraph.}
	\label{FIG:6}
\end{figure*}

As depicted in FIG (\ref{FIG:6}a), we can observe the performance of the HCI-TM algorithm in comparison to other classical algorithms. Specifically, when applied to the US Congressional hypergraph, the HCI-TM algorithm excels by achieving full activation at an early stage. Notably, HC$\rm I_1$ and HC$\rm I_2$ outperform all other algorithms by selecting 75 seed nodes. Among the alternative algorithms, the HHDA algorithm demonstrates the third-best performance, achieving full activation with 102 seed nodes. However, there is still a significant gap compared to the HCI-TM algorithm. On the other hand, the RA algorithm need to select 34.4\% seed nodes (approximately 444 nodes). This striking example effectively highlights the substantial advantages of the HCI-TM algorithm, underscoring its practical utility and application value.

The Vegas-bars-reviews dataset\cite{2020Hypergraph} is derived from the Yelp Kaggle competition data which node represents the user, and hyperedge contains lists of reviewers that reviewed a certain typeof establishment within a month. In FIG (\ref{FIG:6}b), the superiority of the HCI-TM algorithm over other classical algorithms on the Vegas-bars-reviews dataset is clearly demonstrated. As the hypergraph reaches 90\% activation, the HC$\rm I_1$-TM algorithm and HC$\rm I_2$-TM algorithm select 82 and 49 seed nodes respectively, while the HHDA algorithm and NPA algorithm select 124 and 136 seed nodes respectively. Comparing with the HHDA algorithm and NPA algorithm, the HCI algorithm exhibits remarkable advantages, which become more pronounced as the order of the HCI algorithm increases. Among all the algorithms compared, the NP algorithm performs the poorest, requiring 567 seed nodes.

The third real-world hypergraph is the Geometry-question dataset\cite{2020Hypergraph}. The nodes represent the users of MathOverflow and hyperedges are sets of users who answerered a certain question category. FIG (\ref{FIG:6}c) presents the performance of various algorithms on the Geometry-question dataset. The results clearly indicate that the HCI-TM algorithm outperforms all other algorithms. Notably, the HC$\rm I_2$-TM algorithm achieves a hypergraph activation rate of over 90\% by selecting only 20 nodes, which is nearly 45\% higher than the performance of the HHDA algorithm. 

The fourth dataset is Senate-Committees\cite{abh1303}, which nodes represent the members of the United States senate and the hyperedges correspond to committee members. The performance of different algorithms are showed in Fig.(\ref{FIG:6}d), and it outperforms all others. Compared to other classical algorithms, the HC$\rm I_2$-TM algorithm only select 19 seed nodes to achieve the same goal, whereas the HHDA algorithm, which exhibits the best performance among the other classical algorithms, required 36 seeds. That clearly demonstrates the superiority of the HCI-TM algorithm.

The MAG-10 dataset\cite{2020Clustering,2015An} comprises authors as nodes and their corresponding publications as hyperedges within the Microsoft Academic Graph subset. Fig(\ref{FIG:6}e) shows that all algorithms could activate the hypergraph to 0.4 using very few seeds at the initial stage. As propagation process evolved, we observed that the HCI-TM algorithm exhibits a slight advantage over other algorithms. However, HCI-TM algorithm is surpassed by the NPA and HHDA algorithms in the final stage,  but the gap between them was very small. The last dataset we selected was Mathoverflow-answers\cite{2020Minimizing}, where the nodes represent the user, and the hyperedges are the set of questions answered by the users. As can be seen from Fig.(\ref{FIG:6}f), HCI-TM algorithm has obvious advantages compared with other algorithms in the initial stage. However, as the propagation process evolved, the difference between the HCI-TM algorithm and the HHDA and NPA algorithms gradually diminished. In the final stage, NPA algorithm is slightly ahead of HCI-TM algorithm. In addition, we found no first-order phase transitions in the MAG-10 and Mathoverflow-answers datasets when the threshold was set to 0.5. All experiments in real-world hypergraphs show that HCI-TM has strong robustness and superiority.

\section{\label{sec:level1}Conclusion:}

The Influence Maximization problem in hypergraphs has gained attention due to its relevance in real-world scenarios involving high-order interactions. However, this field is still in its early stages, with many heuristic approaches lacking suffcient mathematical analysis and neglecting the dependence of nodes' neighbors. This paper focuses on optimizing node influence based on the threshold models in hypergraphs, but it only solves specific problems. Future research will investigate more general probabilistic agent-based HCI-TM algorithms, applying a message passing theoretical analysis framework to address a broader range of spread and diffusion problems in hypergraphs.

Additionally, the structure of the hypergraphs may have an impact on the occurrence of the cascade phenomenon, which is an area of focus for our future research. Understanding how different hypergraph structures influence the dynamics of information spreading and the cascade phenomenon can provide valuable insights into optimizing Influence Maximization strategies. By further investigating this aspect, we aim to uncover additional factors that contribute to the spread of influence in hypergraphs and develop more effective algorithms for Influence Maximization problem.

\begin{acknowledgments}
We wish to acknowledge the support from the National Key Research and Development Program of China (Grant No.2021ZD0112400), National Natural Science Foundation of China (Grant No.12371516), Liaoning Provincial Natural Science Foundation (Grant No.2022-MS-152), Fundamental Research Funds for the Central Universities(DUT22LAB305).
\end{acknowledgments}

\appendix

\section{Compared Algorithm}

To assess the efficacy of the HCI-TM algorithm, we employed several established algorithms that are commonly utilized in pairwise networks and extended them for application in hypergraphs. Through a comparative analysis with other classical algorithm, we substantiate the superior performance of HCI-TM in efficiently selecting the most influential nodes, thus achieving maximum propagation scale within the hypergraphs.

\subsubsection*{\label{app:subsec}Neighbor Preference Algorithm(NP):}
The NP algorithm employs a sorting technique to prioritize nodes based on their one-layer neighbor count, selecting the node with the highest number of neighbors as the initial seed node for spreading. This process continues until a specific proportion of nodes in the hypergraphs are activated. In order to enhance its efficiency, the NPA algorithm builds upon this approach. After each removal, the neighbors of each node are recalculated, and unactivated nodes with the highest number of neighbors are chosen as seed. Notably, the NPA algorithm avoids selecting already activated nodes as seeds, further optimizing the NP algorithm. The algorithm terminates once the desired proportion of nodes in the hypergraphs are activated.

\subsubsection*{\label{app:subsec}High Hyper Degree Algorithm(HHD):}
The HHD algorithm ranks nodes in descending order based on their hyperdegree and selects the node with the highest hyperdegree as the initial seed node for propagation in the hypergraph. This process continues until a specified proportion of nodes in the hypergraphs are activated. HHDA algorithm is an improvement of HHD algorithm. After each removal, the hyperdegree of each node is recalculated, and the HHDA algorithm prioritizes selecting the remaining unactivated node with the highest hyperdegree as the seed. The algorithm terminates when the desired proportion of nodes in the hypergraphs are activated. By reducing the overlap of already activated nodes as seeds during propagation, the HHDA algorithm enhances the efficiency of the HHD algorithm.

\subsubsection*{\label{app:subsec}PageRank(PR):}
The PageRank algorithm is a widely used method employed by the Google search engine to evaluate the significance of web pages. It operates by analyzing the linkage relationships between web pages, providing a measure of their relative importance in search engine rankings. In our study, we have extended the conventional PageRank algorithm from pairwise networks to hypergraphs,  and the PageRank value of each node in the hypergraph can be defined as follows:

\begin{align}
\left\{ \begin{array}{l}
PR(i) = d(\sum\limits_{j \in \ell i} {\frac{{PR(j)}}{{L(j)}}} ) + \frac{{1 - d}}{N}\\
L(j) = \sum\limits_{{e_\gamma } \in \partial j} {N_\gamma  - 1} 
\end{array} \right.
\label{A1}
\end{align}

In Eq.(\ref{A1}), ${\ell i}$ represents the first layer neighbors of a node, $N_\gamma$ represents the number of nodes in the hyperedge $e_\gamma$, and $d \in (0,1)$ is the damping factor. The PageRank value of each node in the hypergraph is iteratively updated based on the aforementioned equation until convergence is achieved. Once convergence is reached, the nodes are sorted based on their PageRank values. The node with the highest PageRank value is selected as the seed node to activate the hypergraph. The algorithm terminates until a specific proportion of nodes in the hypergraph are activated.

\subsubsection*{\label{app:subsec}Random Algorithm(RA):}
The random algorithm selects seed nodes randomly and propagates them in the hypergraphs. The algorithm terminates until a specific proportion of nodes in the hypergraphs are activated.

\section*{References}

\nocite{*}
\bibliography{aipsamp}

\end{document}